\documentclass [10pt]{article}
\usepackage[left=2cm,top=2.50cm,right=2cm,bottom=2.50cm]{geometry}
\usepackage{mathrsfs}
\usepackage{physics}

\usepackage{amsmath,amssymb,latexsym,color,cancel,graphicx,bbm,colortbl}
\usepackage[english]{babel}
\usepackage[latin1]{inputenc}
\usepackage{ragged2e}
\usepackage{cite}

\begin{document}
\date{}

\title{Berry phase and the Mandel parameter of the non-degenerate parametric amplifier}
\author{J. C. Vega$^{a}$, E. Chore\~no$^{b}$,\\ D. Ojeda-Guill\'en$^{c}$,\footnote{{\it E-mail address:} dojedag@ipn.mx} and R. D. Mota$^{d}$} \maketitle

\begin{minipage}{0.9\textwidth}
\small $^{a}$ Escuela Superior de F{\'i}sica y Matem\'aticas, Instituto Polit\'ecnico Nacional,
Ed. 9, U.P. Adolfo L\'opez Mateos, Alc. Gustavo A. Madero, C.P. 07738 Ciudad de M\'exico, Mexico.\\

\small $^{b}$ Unidad Profesional Interdisciplinaria de Energ\'ia y Movilidad, Instituto Polit\'ecnico Nacional,
Av. Wilfrido Massieu, U.P. Adolfo L\'opez Mateos, Alc. Gustavo A. Madero, C.P. 07738, Ciudad de M\'exico, Mexico.\\

\small $^{c}$ Escuela Superior de C\'omputo, Instituto Polit\'ecnico Nacional,
Av. Juan de Dios B\'atiz esq. Av. Miguel Oth\'on de Mendiz\'abal, Col. Lindavista,
Alc. Gustavo A. Madero, C.P. 07738, Ciudad de M\'exico, Mexico.\\

\small $^{d}$ Escuela Superior de Ingenier{\'i}a Mec\'anica y El\'ectrica, Unidad Culhuac\'an,
Instituto Polit\'ecnico Nacional, Av. Santa Ana No. 1000, Col. San
Francisco Culhuac\'an, Alc. Coyoac\'an, C.P. 04430, Ciudad de M\'exico, Mexico.\\
\end{minipage}

\section*{Abstract}
We study the non-degenerate parametric amplifier problem from an algebraic approach of the $SU(1,1)$ group. We write the Hamiltonian of this problem in terms of the boson generators of the $SU(1,1)$ group and the difference operator. We apply the tilting transformation to our results to exactly solve this Hamiltonian and obtain its energy spectrum and eigenfunctions. Then, by assuming that our Hamiltonian is an explicit function of time we calculate its Berry phase. Finally we obtain the Mandel $Q-$parameter of the photon numbers $n_a$ and $n_b$.

\section{Introduction}

Within quantum optics, the production and study of non-classical states of light has been relevant in recent years, since they allow us to study fundamental concepts within quantum mechanics. Some of the most interesting applications have occurred in interdisciplinary fields such as quantum computing, quantum communication or quantum metrology \cite{Pan,Braun,Reid}.

Among the most efficient and simplest methods to produce non-classical states of light are optical parametric amplifiers, where optical cavities and non-linear crystals are used. In these amplifiers, one photon of a pump field of frequency $\omega_3$ transforms, via the nonlinear medium, into two photons called signal and idler, of frequencies $\omega_1$ and $\omega_2$, respectively. These output beams have the same frequency $(\omega_1=\omega_2)$ and polarization in the degenerate case and different ones $(\omega_1\neq\omega_2)$ in the non-degenerate case \cite{Louisell,Mollow,Mandel}. This pairwise production of photons results in the conservation of the photon-number difference between the signal and idler modes in the absence of any loss. The high correlation between the signal and idler fields is responsible for the generation of a squeezed-vacuum state in the output of the device \cite{Holmes}.

The study of parametric amplifiers and their properties is still relevant today. Some of the most important applications are reported in the Refs. \cite{Good,Renger,Yurke,Yamamoto,Mutus,Grimsmo,Sinha}.

The aim of this work is to study and exactly solve the non-degenerate parametric amplifier. We obtain its eigenfunctions and energy spectrum by using the $su(1,1)$ Lie algebra. Also, we obtain some important properties as its Berry phase \cite{Berry} and the Mandel \cite{Mandel} $Q-$parameter of the photon numbers $n_a$ and $n_b$.

This work is organized as it follows. In Sec. $2$ we write the Hamiltonian of the non-degenerate parametric amplifier in terms of a boson realization of the $su(1,1)$ Lie algebra. Then, we use the $SU(1,1)$ tilting transformation and the eigenfunctions of the two dimensional harmonic oscillator to obtain its energy spectrum and eigenfunctions. The Sec. $3$ is dedicated to compute the Berry phase of the non-degenerate parametric amplifier by supposing that its Hamiltonian is an explicit function of time. The Mandel $Q-$parameter of this problem for the photon numbers $n_a$ and $n_b$ is obtained in Sec. $4$. Finally, we give some concluding remarks.

\section{The non-degenerate parametric amplifier}

The Hamiltonian which describes the stationary non-degenerate parametric amplifier is (with $\hbar = 1$) \cite{Louisell,Walls,Scully}
\begin{equation}
    H = \omega_1 a^{\dagger} a + \omega_2 b^{\dagger} b + \lambda a b + \lambda^{*} a^{\dagger} b^{\dagger}, \label{Hamiltonian}
\end{equation}
where $\lambda = \chi e^{-i \psi}$ is the coupling constant. Now, we can introduce the two-mode realization of the $su(1,1)$ Lie algebra (see Appendix) \cite{Gerrytwomode}
\begin{equation}
    K_0 = \frac{1}{2} (a^{\dagger}a + b^{\dagger}b + 1), \hspace{0.3cm} K_+ = a^{\dagger} b^{\dagger}, \hspace{0.3cm} K_- = ab.\label{Reali}
\end{equation}
These operators together with the operator $J_0 = \frac{1}{2} (a^{\dagger} a - b^{\dagger} b)$, which belongs to the $SU(2)$ group, allow us to write the Hamiltonian in Eq. (\ref{Hamiltonian}) as
\begin{equation}
    H = (\omega_1 + \omega_2) \left( K_0 - \frac{1}{2} \right) +  (\omega_1 - \omega_2) J_0 + \chi e^{-i \psi} K_- + \chi e^{i \psi} K_+.\label{HSU11}
\end{equation}
In order to diagonalize this Hamiltonian we apply the tilting transformation to the stationary Schr\"odinger equation $H\Psi=E\Psi$ as it follows \cite{Gerryberry,Nos1,Nos2}
\begin{equation}
D^{\dagger}(\xi)HD(\xi)D^{\dagger}(\xi)\Psi=ED^{\dagger}(\xi)\Psi.
\end{equation}
Now, we can define the tilted Hamiltonian $H'$ and its wavefunction $\Psi'$ as $H'=D^{\dagger}(\xi)HD(\xi)$ and $\Psi'=D^{\dagger}(\xi)\Psi$, respectively. From the similarity transformations of Eqs. (\ref{DK+})-(\ref{DK0}) we obtain
\begin{equation}\label{Hexp}
    \begin{gathered}
        H' = \left[ (\omega_1 + \omega_2) (2 \beta + 1) + \chi e^{-i \psi} \frac{\xi}{\abs{\xi}} \alpha + \chi e^{i \psi} \frac{\xi^{*}}{\abs{\xi}} \alpha \right] K_0 + \left[ (\omega_1 + \omega_2) \frac{\alpha \xi}{2 \abs{\xi}} + \chi e^{-i \psi} \beta \frac{\xi}{\xi^{*}} + \chi e^{i \psi} (\beta + 1) \right] K_+ \\
        + \left[ (\omega_1 + \omega_2) \frac{\alpha \xi^{*}}{2 \abs{\xi}} + \chi e^{-i \psi} (\beta + 1) + \chi e^{i \psi} \beta \frac{\xi^{*}}{\xi} \right] K_- + (\omega_1 - \omega_2) J_0 - \frac{1}{2} (\omega_1 + \omega_2).
    \end{gathered}
\end{equation}
Therefore, the tilted Hamiltonian becomes diagonal if the coefficients of the operators $K_{\pm}$ vanish. This is achieved by defining the parameters of the coherent state as
\begin{equation}
    \theta = \tanh^{-1}{\left( \frac{2 \chi}{\omega_1 + \omega_2} \right)}, \hspace{0.3cm} \gamma = \psi.\label{para}
\end{equation}
If we substitute these parameters into equation (\ref{Hexp}) we can simplify the tilted Hamiltonian $H'$ to
\begin{equation}
    H' = [(\omega_1 + \omega_2) \cosh{(\theta)} - 2 \chi \sinh{(\theta)}] K_0 + (\omega_1 - \omega_2) J_0 - \frac{1}{2} (\omega_1 + \omega_2).
\end{equation}
By using the results
\begin{equation}\label{coshsinh}
\cosh{(\theta)} = \frac{\omega_1 + \omega_2}{\sqrt{(\omega_1 + \omega_2)^2 - 4 \chi^2}},\quad\quad \sinh{(\theta)} = \frac{2 \chi}{\sqrt{(\omega_1 + \omega_2)^2 - 4 \chi^2}},
\end{equation}
we obtain that the diagonalized Hamiltonian $H'$ becomes
\begin{equation}
    H' = \sqrt{(\omega_1 + \omega_2)^2 - 4\chi^2} K_0 + (\omega_1 - \omega_2) J_0 - \frac{1}{2} (\omega_1 + \omega_2). \label{red}
\end{equation}
Now, let us notice that $2\omega K_0=H_{HO}$, where $H_{HO}$ is the Hamiltonian of the two-dimensional harmonic oscillator. Therefore, since the operators $K_0$ and $J_0$ commute, the eigenfunctions of $H'$ are those of the two-dimensional harmonic oscillator \cite{Schwinger}
\begin{equation}
    \Psi'_{N, m} (r, \varphi) = \frac{1}{\sqrt{\pi}} e^{im \varphi} (-1)^{\frac{N - m}{2}} \sqrt{\frac{2 \left( \frac{N - m}{2} \right)!}{\left( \frac{N + m}{2} \right)!}} r^m L_{\frac{1}{2}(N - m)}^m (r^2) e^{- \frac{1}{2} r^2},
\end{equation}
or
\begin{equation}
    \Psi'_{n_r, m} (r, \varphi) = \frac{1}{\sqrt{\pi}} e^{im \varphi} (-1)^{n_r} \sqrt{\frac{2 \left( n_r \right)!}{\left( n_r + m \right)!}} r^m L_{n_r}^m (r^2) e^{- \frac{1}{2} r^2},
\end{equation}
where $n_r = \frac{1}{2}(N - m)$ is the radial quantum number.

In order to establish the relationship between the quantum numbers of our problem $N, m$ and the quantum numbers of the $SU(1,1)$ group $n, k$, we need to consider the action of the operators $K_0$ and $K^2$ of equations (\ref{k0n}) and (\ref{k2n}) of Appendix on the states $\ket{N,m}$. Thus, from the expression
\begin{equation}
    K^2 \ket{N, m}= \frac{1}{4} (m^2 + 1)\ket{N, m} = k (k - 1) \ket{N, m},
\end{equation}
 we obtain that the Bargmann $k$ index is $k = \frac{1}{2} (m + 1)$. This result of the Bargmann index and the action of $K_0$ on the states $\ket{N,m}$
\begin{equation}
    K_0 \ket{N, m} = \frac{1}{2} (N + 1) \ket{N, m}.\label{K0m}
\end{equation}
lead us to conclude that $n = \frac{1}{2} (N - m) = n_r$. The action of the operator $J_0$ on the states $\ket{N,m}$ can be directly computed from
Eq. (\ref{N-}) or by using the relationship $K^2=J_0^2-\frac{1}{4}$. Hence, we obtain that
\begin{equation}
J_0 \ket{N, m} = \frac{m}{2} \ket{N, m}.\label{N-m}
\end{equation}
Therefore, by substituting the Eqs. (\ref{K0m}) and (\ref{N-m}) into Eq. (\ref{red}) we obtain that the energy spectrum of the non-degenerate parametric amplifier is given by
\begin{equation}
    E = \left[ \sqrt{(\omega_1 + \omega_2)^2 - 4 \chi^2} \left( \frac{1}{2} (N + 1) \right) + \frac{1}{2} (\omega_1 - \omega_2) m - \frac{1}{2} (\omega_1 + \omega_2) \right],\label{es1}
\end{equation}
or in terms of the quantum number $n, m$
\begin{equation}
    E = \left[ \sqrt{(\omega_1 + \omega_2)^2 - 4 \chi^2} \left( n + \frac{m}{2} + \frac{1}{2} \right) + \frac{1}{2} m (\omega_1 - \omega_2) - \frac{1}{2} (\omega_1 + \omega_2) \right].\label{es2}
\end{equation}
The eigenfunctions of the non-degenerate parametric amplifier are obtained from the relationship $\Psi=D(\xi)\Psi'$ and explicitly are the Perelomov number coherent states for the two-dimensional harmonic oscillator \cite{Nos1}
\begin{eqnarray}
\Psi_{n,m}&=&\sqrt{\frac{2\Gamma(n+1)}{\Gamma(n+m+1)}}\frac{(-1)^n}{\sqrt{\pi}}e^{im\varphi}
\frac{(-\zeta^*)^n(1-|\zeta|^2)^{\frac{m}{2}+\frac{1}{2}}(1+\sigma)^n}{(1-\zeta)^{m+1}}\nonumber\\
&&\times e^{-\frac{r^2(\zeta+1)}{2(1-\zeta)}}r^{m}L_n^{m}\left(\frac{r^2\sigma}{(1-\zeta)(1-\sigma)}\right),\label{eigen}
\end{eqnarray}
where $\sigma$ is defined as
\begin{equation}
\sigma=\frac{1-|\zeta|^2}{(1-\zeta)(-\zeta^*)}.
\end{equation}
It is important to note that when we consider the isotropic case $\omega_1 = \omega_2$, the energy spectrum of equations (\ref{es1}) and (\ref{es2}) are properly reduced to those reported in Refs. \cite{Gerryberry,Nos1}. Therefore, we have exactly solved the anisotropic non-degenerate parametric amplifier problem using the tilting transformation and the $SU(1,1)$ group theory.

\section{The Berry phase for the non-degenerate parametric amplifier}

The Berry phase appears when we study the adiabatic evolution of an eigenenergy state in a slowly changing environment and make up a loop in the parameter space \cite{Berry}. In the absence of degeneracy, the Hamiltonian returns to its initial state, and the eigenstate will surely come back to itself when finishing the loop, but there will be a phase difference equal to the time integral of the energy plus the Berry phase. The importance in the study of the Berry phase lies mainly in the fact that it has three key properties: it is gauge invariant, geometrical and has close analogies to gauge field theories and differential geometry \cite{Xiao}.

In order to compute the Berry phase of the non-degenerate parametric amplifier, we can assume that our Hamiltonian of Eq. (\ref{HSU11}) is an explicit function of time
\begin{equation}
H(t)=\Omega(t)\left(K_{0}-\frac{1}{2}\right)+a_{1}(t)K_{+}+a_{2}(t)K_{-}+\Delta\omega(t)J_{0},\label{Ht}
\end{equation}
where $\Omega(t)=\omega_{1}+\omega_{2}$, $\Delta\omega(t)=\omega_{1}-\omega_{2}$ and $a_{2}(t)=a^{*}_{1}(t)$. Here, the coefficient $a_{1}(t)$ is explicitly given by
\begin{equation}
a_{1}(t)=\chi(t)e^{i\psi(t)},
\end{equation}
where $\chi(t)$ and $\psi(t)$ are arbitrary real functions of time.

To describe the quantum dynamics of the system, we shall use the Schr\"odinger picture (with $\hbar=1$)
\begin{equation}
i\frac{d}{dt}|\Psi(t)\rangle=H(t)|\Psi(t)\rangle.\label{Schr}
\end{equation}
The time evolution of the states of our Hamiltonian $H(t)$ can be studied by using the time-dependent nontrivial invariant Hermitian operator $I(t)$ \cite{Lewis1,Lewis2}, which satisfies the condition
\begin{equation}
i\frac{\partial}{\partial t}I(t)+[I(t),H(t)]=0.\label{inavariante}
\end{equation}
By using the $SU(1,1)$ time-dependent displacement operators $D(t)=D(\xi(t))=\exp\{\xi(t)K_{+}-\xi(t)^{*}K_{-}\}$, with $\xi(t)=-\frac{1}{2}\theta(t)e^{-i\gamma(t)}$ and where now $\theta(t)$ and $\gamma(t)$ are arbitrary real functions of time, we can define the invariant operator $I(t)$ as (see Ref. \cite{Lai})
\begin{equation}
I(t)=D(t)K_{0}D^{\dag}(t).\label{id}
\end{equation}
From the definition of the displacement operator we obtain that $D^{\dag}(\xi)=D(-\xi)$. Thus, from equation (\ref{DK0}) of Appendix we obtain that the invariant operator $I(t)$ is explicitly
\begin{equation}
I(t)=\cosh(\theta)K_{0}+\frac{\sinh(\theta)}{2}e^{-i\gamma}K_{+}+\frac{\sinh(\theta)}{2}e^{i\gamma}K_{-}.\label{I1}
\end{equation}
Now, if we substitute the time-dependent Hamiltonian (\ref{Ht}) and the invariant operator $I(t)$ of expression (\ref{I1}) into the equation (\ref{inavariante}), we obtain that the time-dependent physical parameters $\Omega(t)$, $\Delta\omega(t)$, $\chi(t)$ and $\psi(t)$ are related with the parameters $\theta(t)$ and $\gamma(t)$ as follows
\begin{equation}
\dot{\theta}=-2\chi\sin(\psi+\gamma),\quad\quad(\dot{\gamma}-\Omega)\sinh(\theta)=-2\chi\cosh(\theta)\cos(\psi+\gamma).\label{cond1}
\end{equation}

In addition, by using of the BCH identity (Eq. (\ref{BCH}) of Appendix) the operator $i\frac{\partial}{\partial t}$ is transformed under the $SU(1,1)$ time-dependent displacement operators $D(\xi)$ as
\begin{equation}
D^{\dag}(t)\left(i\frac{\partial}{\partial t}\right)D(t)=\dot{\gamma}(\cosh(\theta)-1)K_{0}-\frac{e^{-i\gamma}}{2}\left(\dot{\gamma}\sinh(\theta)+i\dot{\theta}\right)K_{+}-\frac{e^{i\gamma}}{2}\left(\dot{\gamma}\sinh(\theta)-i\dot{\theta}\right)K_{-}.\label{dt1}
\end{equation}

As it is shown in Ref. \cite{Lewis2}, if the eigenstates of the invariant operator satisfy the Schr\"odinger equation its eigenvalues are real. Therefore, from the Eq. (\ref{k0n}) of Appendix and the expression (\ref{id}) we have
\begin{equation}
D(t)K_{0}|k,n\rangle=(k+n)D(t)|k,n\rangle,\quad\quad I(t)D(t)|k,n\rangle=(k+n)D(t)|k,n\rangle.
\end{equation}
Thus, the eigenvalues of the invariant operator $I(t)$ are $(k+n)$ and its eigenstates are
$D(t)|k,n\rangle=|\zeta(t),k,n\rangle$, which are the $SU(1,1)$ Perelomov number coherent states of Eq. (\ref{PNCS}),
but now these are functions of time.

Moreover, if the states $|\Psi(t)\rangle$ satisfy the Eq. (\ref{Schr}), then they can be expanded through the states $|\zeta(t),k,n\rangle$ in the form
\begin{equation}
|\Psi(t)\rangle=\sum_{n}a_{n}e^{i\alpha_{n}}|\zeta(t),k,n\rangle,
\end{equation}
where according to Lewis \cite{Lewis2} the phase $\alpha_n$ is given as
\begin{equation}
\alpha_n=\int_{0}^{t}dt'\langle \lambda,\kappa|i\frac{\partial}{\partial t'}-H(t')|\lambda,\kappa\rangle.\label{T-phase}
\end{equation}
Here, $|\lambda,\kappa\rangle$ are the eigenstates and $\lambda$ are the eigenvalues of the invariant operator $I(t)$.
For our problem $|\lambda,\kappa\rangle=|\zeta(t),k,n\rangle=D(\xi)|N,m\rangle$. Therefore, by substituting the equation (\ref{dt1}) and the time-dependent results of expressions (\ref{DK+})-(\ref{DK0}) of Appendix into equation (\ref{T-phase}) we obtain that
the phase $\alpha_n$ in a non-adiabatic process is given by
\begin{align}\nonumber
\alpha_{n}&=\int_{0}^{t}\langle N,m|\left[(\dot{\gamma}-\Omega)(\cosh(\theta)-1)+2\chi\cos(\gamma+\psi)\sinh(\theta)-\Omega\right]K_0-\Delta\omega J_0+\frac{1}{2}\Omega |N,m\rangle dt',\\
&=\int_{0}^{t}\left[\frac{1}{2}(N+1)\left[(\dot{\gamma}-\Omega)(\cosh(\theta)-1)+2\chi\cos(\gamma+\psi)\sinh(\theta)-\Omega\right]-\Delta\omega\frac{m}{2}+\frac{1}{2}\Omega\right]dt'.\label{nophase1}
\end{align}
In this expression we have used the results of Eqs. (\ref{K0m}) and (\ref{N-m}). Unlike in a non-adiabatic process, in an adiabatic process we have that $\dot{\theta}=\dot{\gamma}=0$ and the equations (\ref{cond1}) become
\begin{equation}
\psi +\gamma=n\pi,\quad\quad\tanh(\theta)=\frac{2\chi}{\Omega}(-1)^{n}.\label{1cond1}
\end{equation}
If we set $n=0$, the above conditions are reduced to the time-dependent versions of the expressions
\begin{equation}
\psi=-\gamma,\quad\quad\tanh(\theta)=\frac{2\chi}{\omega_{1}+\omega_{2}}.
\end{equation}
We note that these results are very similar to those obtained in Section 2 for the diagonalization process of the time-independent parametric amplifier Hamiltonian (see equation (\ref{para})). Moreover, from equations (\ref{DK+})-(\ref{DK0}), it can be shown that
\begin{equation}
D^{\dag}(t)HD(t)=\left[\Omega\cosh{(\theta)}-2\chi\sinh{(\theta)}\right]K_0+\Delta\omega J_0-\frac{1}{2}\Omega.
\end{equation}
Therefore, from the results of expressions (\ref{K0m}) and (\ref{N-m}) it follows that
\begin{equation}
D(t)|N,m\rangle=\left\{\left[\Omega(t)\cosh{(\theta(t))}-2\chi(t)\sinh{(\theta(t))}\right]\left(\frac{N+1}{2}\right)+\Delta\omega(t) \left(\frac{m}{2}\right)-\frac{1}{2}\Omega(t)\right\}|N,m\rangle.
\end{equation}
This shows that, in the adiabatic limit, $D(t)|N,m\rangle$ becomes an instantaneous eigenstate of the Hamiltonian $H(t)$ and the invariant operator $I(t)$, which is a crucial requirement for the computation of the Berry phase \cite{Lai}.

In an adiabatic process the phase of the states $D(\xi)|N,m\rangle$ is reduced to
\begin{equation}
\alpha_{n}=-\int_{0}^{t}\left[\frac{1}{2}(N+1)\sqrt{\Omega(t')^{2}-4\chi^{2}(t')}+\Delta\omega(t')\frac{m}{2}-\frac{1}{2}\Omega\right]dt'.\label{dphase1}
\end{equation}

The second term of equation (\ref{T-phase}) is known as the dynamical phase and is defined as
\begin{equation}
\dot{\epsilon}_{N}=\langle N,m|D^{\dag}(\xi)H(t')D(\xi)|N,m\rangle\label{dyphase},
\end{equation}
while the first term is known as the Berry phase
\begin{equation}
\dot{\Gamma}_{N}=i\langle N,m|D^{\dag}(\xi)\frac{\partial}{\partial t}D(\xi)|N,m\rangle.\label{berphase}
\end{equation}
Thus, the Berry phase of the states $D(\xi)|N,m\rangle$ is obtained in the adiabatic limit as follows
\begin{equation}
\Gamma_{N}(T)=\frac{1}{2}(N+1)\oint (\cosh(\theta)-1)d\psi,\label{Berrysu11}
\end{equation}
where $T$ denotes the period. It is obvious that in this case the Berry phase do not depend on an explicit form of $\psi(t)$. Therefore, since $\oint d\psi=2\pi$ for one period, we finally obtain that the Berry phase for the non-degenerate parametric amplifier only depends on the time-dependent parameters in the Hamiltonian, i.e.,
\begin{equation}
\Gamma_{N}(T)=\pi\left(N+1\right)\frac{\Omega(t)-\sqrt{\Omega^{2}(t)-4\chi^{2}(t)}}{\sqrt{\Omega^{2}(t)-4\chi^{2}(t)}}.
\end{equation}

\section{The Mandel $Q-$parameter of the non-degenerate parametric amplifier}

Since the appearance of the first experiments of light emission, the study of statistical properties and intensity correlations have been key in their study. Among all the parameters introduced to study and characterize statistical properties, the Mandel parameter is one of the most used \cite{Mandel,Mandel1}. This parameter allows us to measure the deviation from Poisson distribution, in order to distinguish quantum processes from classical ones.
The Mandel $Q-$parameter is defined by
\begin{equation}
Q=\frac{\langle\hat{n}^{2}\rangle-\langle\hat{n}\rangle^{2}}{\langle\hat{n}\rangle}-1,\label{Mandel}
\end{equation}
and where
\begin{equation}
Q=\left\{ \begin{array}{lcc}
             >0, &   super~Poissonian~distribution  &  \\
             \\ =0, &  Poissonian~distribution~(coherent~state)  &  \\
             \\ <0, &  sub-Poissonian  & \\
             \\ =-1,&  number~state.
             \end{array}
   \right.
\end{equation}

In this Section we will compute the Mandel $Q$-parameter for the eigenstates $|\Psi\rangle=D(\xi)|\Psi'\rangle$ of the non-degenerate parametric amplifier. Hence, we can obtain the Mandel $Q_a$ parameter from the expression (\ref{Mandel}) as
\begin{equation}
Q_{a}=\frac{\langle n^{2}_{a}\rangle-\langle n_{a}\rangle^{2}}{\langle n_{a}\rangle}-1.\label{Qa1}
\end{equation}
In this expression we have that
\begin{equation}\label{n2a}
\langle n_{a}^{2}\rangle=\langle(a^{\dag}a)^{2}\rangle=\langle\Psi'|\left[ D^{\dag}(\xi)a^{\dag}aD(\xi)\right]^{2}|\Psi'\rangle,
\end{equation}
\begin{equation}\label{na2}
\langle n_{a}\rangle^{2}=\langle a^{\dag}a\rangle^{2}=\left[\langle\Psi'|D^{\dag}(\xi)a^{\dag}aD(\xi)|\Psi'\rangle\right]^{2},
\end{equation}
and
\begin{equation}\label{na}
\langle n_{a}\rangle=\langle a^{\dag}a\rangle=\langle\Psi'|D^{\dag}(\xi)a^{\dag}aD(\xi)|\Psi'\rangle,
\end{equation}
where $|\Psi'\rangle=|N,m\rangle=|\rangle_{N,m}$. Now, we can express the term $a^{\dag}a$ as a linear combination of the operators $K_0$ and $J_0$ as
\begin{equation}\label{a+a}
a^{\dag}a=K_{0}+J_{0}-\frac{1}{2}.
\end{equation}

By using the $SU(1,1)$ displacement operator $D(\xi)$ of Eq. (\ref{do}) and the result of expression (\ref{a+a}) we arrive to
\begin{equation}\label{daad}
D^{\dag}(\xi)a^{\dag}aD(\xi)=\cosh(\theta)K_{0}-\frac{e^{-i\gamma}}{2}\sinh(\theta)K_{+}-\frac{e^{i\gamma}}{2}\sinh(\theta)K_{-}+J_{0}-\frac{1}{2}.
\end{equation}
Thus, by substituting Eq. (\ref{daad}) into Eq. (\ref{n2a}) we have
\begin{equation}
\langle n^{2}_{a}\rangle=\cosh^{2}(\theta)\langle K^{2}_{0}\rangle_{N,m}+2\cosh(\theta)\langle J_{0}K_{0}-K_{0}/2\rangle_{N,m}+\frac{\sinh^{2}(\theta)}{4}\left[\langle K_{+}K_{-}\rangle_{N,m}+\langle K_{-}K_{+}\rangle_{N,m}\right]+\langle\left( J_{0}-1/2\right)^{2}\rangle_{N,m}.\label{n2aff}
\end{equation}
Similarly, from Eqs. (\ref{daad}), (\ref{na2}) and (\ref{na}) it is easily seen that
\begin{align}
\langle n_{a}\rangle^{2}=\cosh^{2}(\theta)\langle K_{0}\rangle^{2}_{N,m}+\langle J_{0}\rangle^{2}_{N,m}+2\cosh(\theta)\langle K_{0}\rangle_{N,m}\langle J_{0}\rangle_{N,m}-\cosh(\theta)\langle K_{0}\rangle_{N,m}-\langle J_{0}\rangle_{N,m}+\frac{1}{4},\label{na2ff}
\end{align}
and
\begin{align}
\langle n_{a}\rangle=\cosh(\theta)\langle K_{0}\rangle_{N,m}+\langle J_{0}\rangle_{N,m}-\frac{1}{2}.\label{naff}
\end{align}
Therefore, from the results of equations (\ref{n2aff})-(\ref{naff}), the Mandel $Q_a$ parameter can be expressed in terms of the $su(1,1)$ Lie algebra generators as
\begin{equation}
Q_{a}=\frac{\sinh^{2}(\theta)}{4}\left[\frac{\langle K_{+}K_{-}\rangle_{N,m}+\langle K_{-}K_{+}\rangle_{N,m}}{\cosh(\theta)\langle K_{0}\rangle_{N,m}+\langle J_{0}\rangle_{N,m}-\frac{1}{2}}\right]-1.
\end{equation}
We can compute the expected values in this expression from Eq. (\ref{Reali}), and the actions of the boson operators $a$ and $b$ on the states $|N,m\rangle$ (see Eqs. (\ref{acta}) and (\ref{actb}))
\begin{equation}
\langle K_{0}\rangle_{N,m}=\frac{N+1}{2},\quad\quad\langle J_{0}\rangle_{N,m}=\frac{m}{2},\quad\quad\langle K_{+}K_{-}\rangle_{N,m}=\frac{N^{2}-m^{2}}{4},\quad\quad \langle K_{-}K_{+}\rangle_{N,m}=\frac{N^{2}-m^{2}}{4}+N+1.\label{12}
\end{equation}
Thus, in terms of the quantum numbers $N,m$ it follows that
\begin{equation}
Q_{a}=\frac{\sinh^{2}(\theta)}{4}\left[\frac{N^{2}-m^{2}+2N+2}{\cosh(\theta)(N+1)+m-1}\right]-1.\label{Q_a}
\end{equation}
Furthermore, by using the results of Eq. (\ref{coshsinh}) with $\Omega=\omega_{1}+\omega_{2}$, we obtain that the $Q_a$-parameter for the non-degenerate parametric amplifier, the expression (\ref{Q_a}) can be rewritten as
\begin{equation}
Q_{a}=\frac{\chi^2}{\sqrt{\Omega^{2}-4\chi^{2}}}\left[\frac{N^{2}-m^{2}+2N+2}{\Omega(N+1)+(m-1)\sqrt{\Omega^{2}-4\chi^{2}}}\right]-1.
\end{equation}

Analogously, we can compute the Mandel $Q_{b}$ parameter considering that
\begin{equation}
b^{\dag}b=K_{0}-J_{0}-\frac{1}{2},
\end{equation}
and therefore we have the following results for $\langle n^{2}_{b}\rangle$, $\langle n_{b}\rangle^{2}$ and $\langle n_{b}\rangle$
\begin{equation}
\langle n^{2}_{b}\rangle=\cosh^{2}(\theta)\langle K^{2}_{0}\rangle_{N,m}-2\cosh(\theta)\langle J_{0}K_{0}+K_{0}/2\rangle_{N,m}+\frac{\sinh^{2}(\theta)}{4}\left[\langle K_{+}K_{-}\rangle_{N,m}+\langle K_{-}K_{+}\rangle_{N,m}\right]+\langle\left( J_{0}+1/2\right)^{2}\rangle_{N,m},
\end{equation}
\begin{align}
\langle n_{b}\rangle^{2}=\cosh^{2}(\theta)\langle K_{0}\rangle^{2}_{N,m}+\langle J_{0}\rangle^{2}_{N,m}-2\cosh(\theta)\langle K_{0}\rangle_{N,m}\langle J_{0}\rangle_{N,m}-\cosh(\theta)\langle K_{0}\rangle_{N,m}+\langle J_{0}\rangle_{N,m}+\frac{1}{4},
\end{align}
\begin{align}
\langle n_{b}\rangle=\cosh(\theta)\langle K_{0}\rangle_{N,m}-\langle J_{0}\rangle_{N,m}-\frac{1}{2}.
\end{align}
Thus, by substituting these results in the definition of the Mandel parameter (Eq. (\ref{Mandel})) we obtain
\begin{equation}
Q_{b}=\frac{\sinh^{2}(\theta)}{4}\left[\frac{\langle K_{+}K_{-}\rangle_{N,m}+\langle K_{-}K_{+}\rangle_{N,m}}{\cosh(\theta)\langle K_{0}\rangle_{N,m}-\langle J_{0}\rangle_{N,m}-\frac{1}{2}}\right]-1.
\end{equation}
Finally by using the relations (\ref{coshsinh}) and (\ref{12}), we obtain that the Mandel parameter $Q_b$ for the non-degenerate parametric amplifier is explicitly given by
\begin{equation}
Q_{b}=\frac{\chi^2}{\sqrt{\Omega^{2}-4\chi^{2}}}\left[\frac{N^{2}-m^{2}+2N+2}{\Omega(N+1)-(m+1)\sqrt{\Omega^{2}-4\chi^{2}}}\right]-1.
\end{equation}

\section{Concluding remarks}

In this paper we were able to obtain the exact solution of the non-degenerate parametric amplifier by means of the tilting transformation and the unitary irreducible representation theory of the $SU(1,1)$ group. This allowed us to diagonalize our original Hamiltonian and write it in terms of the Hamiltonian of the two-dimensional harmonic oscillator.

Furthermore, by assuming that our Hamiltonian was explicitly time-dependent and using the similarity transformations of the operator $i\frac{\partial}{\partial t}$, we computed the Berry phase of the non-degenerate parametric amplifier.

Finally, the bosonic realization of the $su(1,1)$ Lie algebra and the similarity transformations of the displacement operator $D(\xi)$ allowed us to calculate the Mandel parameters of the photon numbers $n_a$ and $n_b$.

\section*{Acknowledgments}

This work was partially supported by SNII-M\'exico, EDI-IPN, SIP-IPN Project Number $20230633$.\\
We appreciate the observations made by the anonymous referee to improve our work.

\section*{Disclosures}

The authors declare no conflicts of interest.

\section*{Data Availability}

No data were generated or analyzed in the presented research.

\section{Appendix.}

\subsection{The $su(1,1)$ Lie algebra and its coherent states}

Three operators $K_{\pm}, K_0$ close the $su(1,1)$ Lie algebra if they satisfy the commutation relations \cite{Vourdas}
\begin{eqnarray}
[K_{0},K_{\pm}]=\pm K_{\pm},\quad\quad [K_{-},K_{+}]=2K_{0}.\label{com}
\end{eqnarray}
The Casimir operator $K^{2}$, which commutes with these three operators is given for any irreducible representation of this group by
\begin{equation}
K^2=K^2_0-\frac{1}{2}\left(K_+K_-+K_-K_+\right).
\end{equation}
The action of these four operators on the Fock space states
$\{|k,n\rangle, n=0,1,2,...\}$ is
\begin{equation}
K_{+}|k,n\rangle=\sqrt{(n+1)(2k+n)}|k,n+1\rangle,\label{k+n}
\end{equation}
\begin{equation}
K_{-}|k,n\rangle=\sqrt{n(2k+n-1)}|k,n-1\rangle,\label{k-n}
\end{equation}
\begin{equation}
K_{0}|k,n\rangle=(k+n)|k,n\rangle,\label{k0n}
\end{equation}
\begin{equation}
K^2|k,n\rangle=k(k-1)|k,n\rangle.\label{k2n}
\end{equation}
The Bargmann index $k$ determines a representation of the $su(1,1)$ algebra. The discrete series are those for which $k>0$.

The displacement operator $D(\xi)$ is defined in terms of the creation and annihilation operators $K_+, K_-$ as
\begin{equation}
D(\xi)=\exp(\xi K_{+}-\xi^{*}K_{-}),\label{do}
\end{equation}
where $\xi=-\frac{1}{2}\theta e^{-i\gamma}$, $-\infty<\theta<\infty$ and $0\leq\gamma\leq2\pi$.
The so-called normal form of the displacement operator is given by
\begin{equation}
D(\xi)=\exp(\zeta K_{+})\exp(\eta K_{0})\exp(-\zeta^*K_{-})\label{normal},
\end{equation}
where  $\zeta=-\tanh(\frac{1}{2}\theta)e^{-i\gamma}$ and $\eta=-2\ln \cosh|\xi|=\ln(1-|\zeta|^2)$ \cite{Gerry}.

The $SU(1,1)$ Perelomov coherent states are defined as the action of the displacement operator $D(\xi)$
on the lowest normalized state $|k,0\rangle$ as \cite{Perellibro}
\begin{equation}
|\zeta\rangle=D(\xi)|k,0\rangle=(1-|\zeta|^2)^k\sum_{n=0}^\infty\sqrt{\frac{\Gamma(n+2k)}{n!\Gamma(2k)}}\zeta^n|k,n\rangle.\label{PCN}
\end{equation}
The $SU(1,1)$ Perelomov number coherent state $|\zeta,k,n\rangle$ is defined as the action of the displacement operator $D(\xi)$ on an arbitrary
excited state $|k,n\rangle$ \cite{Nos1,Nos3}
\begin{eqnarray}
|\zeta,k,n\rangle &=&\sum_{s=0}^\infty\frac{\zeta^s}{s!}\sum_{j=0}^n\frac{(-\zeta^*)^j}{j!}e^{\eta(k+n-j)}
\frac{\sqrt{\Gamma(2k+n)\Gamma(2k+n-j+s)}}{\Gamma(2k+n-j)}\nonumber\\
&&\times\frac{\sqrt{\Gamma(n+1)\Gamma(n-j+s+1)}}{\Gamma(n-j+1)}|k,n-j+s\rangle.\label{PNCS}
\end{eqnarray}

The similarity transformations $D^{\dag}(\xi)K_{+}D(\xi)$, $D^{\dag}(\xi)K_{-}D(\xi)$, and
$D^{\dag}(\xi)K_{0}D(\xi)$ of the $su(1,1)$ Lie algebra generators are computed by using the displacement operator $D(\xi)$ an the Baker-Campbell-Hausdorff identity
\begin{equation}\label{BCH}
e^{A}Be^{-A}=B+[A,B]+\frac{1}{2!}[A,[A,B]]+\frac{1}{3!}[A,[A,[A,B]]]+...
\end{equation}
These results explicitly are
\begin{equation}
D^{\dag}(\xi)K_{+}D(\xi)=\frac{\xi^{*}}{|\xi|}\alpha K_{0}+\beta\left(K_{+}+\frac{\xi^{*}}{\xi}K_{-}\right)+K_{+},\label{DK+}
\end{equation}
\begin{equation}
D^{\dag}(\xi)K_{-}D(\xi)=\frac{\xi}{|\xi|}\alpha K_{0}+\beta\left(K_{-}+\frac{\xi}{\xi^{*}}K_{+}\right)+K_{-},\label{DK-}
\end{equation}
\begin{equation}
D^{\dag}(\xi)K_{0}D(\xi)=(2\beta+1)K_{0}+\frac{\alpha\xi}{2|\xi|}K_{+}+\frac{\alpha\xi^{*}}{2|\xi|}K_{-},\label{DK0}
\end{equation}
where $\alpha=\sinh(2|\xi|)$ and $\beta=\frac{1}{2}\left[\cosh(2|\xi|)-1\right]$.

\subsection{The $su(1,1)$ Jordan-Schwinger realization}

A particular realization of the $su(1,1)$ Lie algebra is given by the Jordan-Schwinger operators \cite{Gerrytwomode}
\begin{equation}
K_0=\frac{1}{2}\left(a^{\dag}a+b^{\dag}b+1\right), \quad K_+=a^{\dag}b^{\dag},\  \quad K_-= ba,\label{su11}
\end{equation}
where the two sets of operators $(a,a^{\dag})$ and $(b,b^{\dag})$ satisfy the bosonic algebra
\begin{equation}
[a,a^{\dag}]=[b,b^{\dag}]=1, \quad\quad[a,b^{\dag}]=[a,b]=0.
\end{equation}
If we introduce the operator $J_0$ defined as half the difference of the number operators of the two oscillators, then $J_0$ commutes with all
the generators of this algebra realization. Hence, we have the following results\cite{Vourdasanalytic}
\begin{equation}
K^2=J_0^2-\frac{1}{4}, \quad\quad J_0=\frac{1}{2}\left(a^{\dag}a-b^{\dag}b\right),\nonumber
\end{equation}
\begin{equation}
[J_0,K_0]=[J_0,K_+]=[J_0,K_-]=0.
\end{equation}
In fact, the operator $J_0$ is one of the generators of the $su(2)$ Jordan-Schwinger Lie algebra realization.

On the other hand, in polar coordinates the creation and annihilation operators take the form
\begin{eqnarray}
a=\frac{e^{-i\varphi}}{2}\left[r+\frac{\partial}{\partial r}-\frac{i}{r}\frac{\partial}{\partial \varphi}\right],\quad\quad
a^{\dag}=\frac{e^{i\varphi}}{2}\left[r-\frac{\partial}{\partial r}-\frac{i}{r}\frac{\partial}{\partial \varphi}\right],\label{apolar}
\end{eqnarray}
\begin{eqnarray}
b=\frac{e^{i\varphi}}{2}\left[r+\frac{\partial}{\partial r}+\frac{i}{r}\frac{\partial}{\partial \varphi}\right],\quad\quad
b^{\dag}=\frac{e^{-i\varphi}}{2}\left[r-\frac{\partial}{\partial r}+\frac{i}{r}\frac{\partial}{\partial \varphi}\right].\label{bpolar}
\end{eqnarray}
The action of these operators on the eigenfunctions of the two-dimensional harmonic oscillator $|N,m\rangle$ is given by \cite{Wallace}
\begin{eqnarray}
a|N,m\rangle=\sqrt{\frac{1}{2}(N+m)}|N-1,m-1\rangle,\quad a^{\dag}|N,m\rangle=\sqrt{\frac{1}{2}(N+m)+1}|N+1,m+1\rangle,\label{acta}
\end{eqnarray}
\begin{eqnarray}
b|N,m\rangle=\sqrt{\frac{1}{2}(N-m)}|N-1,m+1\rangle,\quad b^{\dag}|N,m\rangle=\sqrt{\frac{1}{2}(N-m)+1}|N+1,m-1\rangle.\label{actb}
\end{eqnarray}
Thus, by substituting equations (\ref{apolar}) and (\ref{bpolar}) into equation (\ref{Reali}), we obtain that the $su(1,1)$ Lie algebra generators $K_{\pm}, K_0$ and the Casimir operator $K^2$ in polar coordinates are given by \cite{Nos1}
\begin{equation}
    \begin{gathered}
        K_+ = \frac{1}{4} \left[ r^2 - 2r \frac{\partial}{\partial r} - 2 + \frac{\partial^2}{\partial r^2} + \frac{1}{r} \frac{\partial}{\partial r} + \frac{1}{r^2} \frac{\partial^2}{\partial \varphi^2} \right], \\
        K_- = \frac{1}{4} \left[ r^2 + 2r \frac{\partial}{\partial r} + 2 + \frac{\partial^2}{\partial r^2} + \frac{1}{r} \frac{\partial}{\partial r} + \frac{1}{r^2} \frac{\partial^2}{\partial \varphi^2} \right], \\
        K_0 = \frac{1}{4} \left[ r^2 - \frac{\partial^2}{\partial r^2} - \frac{1}{r} \frac{\partial}{\partial r} + \frac{1}{r^2} \frac{\partial^2}{\partial \varphi^2} \right], \\
        K^2 = -\frac{1}{4} \left( 1 + \frac{\partial^2}{\partial \varphi^2} \right).
    \end{gathered}
\end{equation}
Similarly we can obtain that the operator $J_0$ is written in polar coordinates by
\begin{equation}
    J_0 = \frac{i}{2} \frac{\partial}{\partial \varphi}.\label{N-}
\end{equation}

\end{document}